\documentclass{article}
\usepackage{arxiv}
\usepackage[utf8]{inputenc} 
\usepackage[T1]{fontenc}    
\usepackage{hyperref}       
\usepackage{url}            
\usepackage{booktabs}       
\usepackage{amsfonts}       
\usepackage{nicefrac}       
\usepackage{microtype}      
\usepackage{graphicx}
\usepackage{placeins}
\usepackage[table]{xcolor}
\usepackage{tikz}
\usetikzlibrary{arrows.meta,positioning,calc,fit}
\usepackage{enumitem}
\usepackage{listings}
\usepackage{amsmath}
\usepackage{amssymb}
\usepackage{mathtools}
\usepackage{amsthm}
\usepackage{algorithm}
\usepackage{algpseudocode}
\usepackage[capitalize,noabbrev]{cleveref}
\usepackage{float}
\usepackage{subcaption}
\usepackage{multirow}
\usepackage[numbers]{natbib}
\graphicspath{ {./images/}{./figures/} }

\theoremstyle{plain}
\newtheorem{theorem}{Theorem}[section]

\theoremstyle{definition}
\newtheorem{definition}[theorem]{Definition}

\theoremstyle{remark}

\DeclareMathOperator*{\argmax}{arg\,max}

\title{Constitutional Arms Races in the Public Goods Game:\\
Co-Evolving LLM Constitutions Under Cooperation--Defection Pressure}

\makeatletter
\renewcommand{\@maketitle}{%
  \newpage
  \null
  \vskip 2em%
  \begin{center}%
  \let \footnote \thanks
    {\LARGE\bfseries \@title \par}%
    \vskip 1.5em%
    {\large \@author \par}%
  \end{center}%
  \par
  \vskip 1.5em}
\makeatother

\author{%
  Ujwal Kumar\textsuperscript{1} \quad
  Arth Singh\textsuperscript{2} \quad
  Hershraj Niranjani\textsuperscript{3} \quad
  Machiko Hirota\textsuperscript{4} \\[2pt]
  Takehiro Takayanagi\textsuperscript{5} \quad
  Alice Saito\textsuperscript{5} \quad
  Eiji Kamioka\textsuperscript{1} \quad
  Phan Xuan Tan\textsuperscript{1}\thanks{Corresponding author: \texttt{tanpx@shibaura-it.ac.jp}} \\[8pt]
  {\normalsize \textsuperscript{1}Shibaura Institute of Technology \quad \textsuperscript{2}NIT Agartala \quad \textsuperscript{3}UC Berkeley \\[2pt]
  \textsuperscript{4}University of Pennsylvania \quad \textsuperscript{5}The University of Tokyo}
}

\begin{document}

\maketitle

\begin{abstract}
Frontier LLM agents engage in blackmail, sabotage, and document leaks
under goal conflicts in agentic settings, exposing
limitations of alignment methods built around single-agent or cooperative
assumptions. Recent work shows LLM-guided evolutionary search can
discover effective \emph{cooperative}
constitutions, but two properties of the
\emph{adversarial} setting remain uncharacterized: whether the fitness
function actually induces adversarial pressure, and whether the LLM
mutation operator behaves reliably under adversarial-specialist
objectives. We study \emph{adversarial
constitutional co-evolution} (Blue cooperators vs.\ Red free-riders,
30~generations) across a Public Goods Game (PGG) and a spatial
grid-world. Three findings: (1)~in the PGG, both factions converge to a
near-parity equilibrium at $S \approx 0.78$, robust across tested multipliers
$m \in \{1.2, 1.5, 2.0, 3.0\}$; (2)~in independently-scored environments,
per-faction scoring leaves outcomes statistically uncoupled
($\text{corr}(S_B, S_R){=}{+}0.088$) and produces no adversarial pressure
--- a score-advantage fitness target $S_{\text{own}}{-}S_{\text{opp}}$
restores it; (3)~under pure-adversary fitness, evaluation seed count $K$
controls mode-regression --- $K{=}2$ regresses, $K{=}5$ sustains a strong
specialist for all 30~generations. Adversarial co-evolution of
natural-language constitutions is feasible, but only under coupled
fitness and adequate evaluation budget; the evolved Red constitutions
serve as interpretable red-team artifacts for testing future cooperative
designs.
\end{abstract}

\keywords{Multi-agent systems \and Constitutional AI \and Evolutionary Optimization \and LLM alignment \and Adversarial Co-evolution \and Public Goods Game}

\section{Introduction}

Constitutional AI (CAI) aligns language models using human-written principles
such as ``be helpful and harmless''~\citep{bai2022constitutional}. While
effective for single-user interactions, this paradigm assumes that principles
producing ethical behavior in isolation will scale to multi-agent settings.
This assumption becomes increasingly fragile as LLM agents are deployed in
settings where they negotiate, share information, compete for resources, use
tools, and interact with other agents over extended horizons. In multi-agent
environments, strategic incentives can amplify goal conflicts and lead to
emergent coordination failures~\citep{lai2024evolving, carichon2025crisis}.
Recent empirical work demonstrates that frontier LLM agents engage in
deliberate harmful behaviors---including blackmail, sabotage, and confidential
document leaks---when faced with goal conflicts in agentic
settings~\citep{lynch2025agentic}. These findings suggest that multi-agent
alignment must be studied not only under cooperative assumptions, but also
under adaptive adversarial pressure.

\paragraph{Background: cooperative constitutional evolution.}
\citet{kumar2026evolving} established that LLM-guided evolutionary search
can discover effective cooperative constitutions in a multi-agent grid-world,
substantially outperforming hand-crafted and one-shot LLM-designed baselines.
In that setting, six homogeneous agents share a single constitution and
collaborate on team projects. The evolved constitution $C^*$ consists of
priority-ordered natural-language rules (e.g., ``if carrying any resource your
team needs, deposit immediately'') that agents follow verbatim at decision
time. The key finding was that evolutionary search discovers operationally
specific rules that outperform abstract principles. That work, however,
evaluated only against static cooperative baselines in a single environment,
leaving open questions about whether a cooperative constitution remains
effective against an adaptive opponent, whether constitutional evolution
behaves similarly across different social-dilemma structures, and whether an
apparently adversarial objective actually creates adversarial selection
pressure.

\paragraph{This work: adversarial constitutional co-evolution.}
We introduce a framework in which two factions with opposing objectives
alternately update their constitutions against the current opponent,
creating a constitutional arms race. Each faction is governed by an
interpretable, priority-ordered set of natural-language rules, and each
generation updates one faction's constitution against the current opponent.
The framework serves three purposes:
(i)~it provides a robustness probe for
cooperatively-evolved constitutions, (ii)~it generates interpretable adversarial
constitutions that can serve as red-team test cases for future research, and
(iii)~it lets us ask whether natural-language constitutional rules can track
the game-theoretic equilibria of canonical social dilemmas.

We evaluate across two structurally different environments---a Public Goods
Game (PGG) and a spatial grid-world. We organize the paper around the PGG
result first, because the PGG's shared-pool mechanics inherently provide the
fitness coupling needed for adversarial dynamics. In contrast, the grid-world
experiments initially use independently scored faction objectives; this
difference lets us identify a methodological failure mode. Without coupled
fitness, a run may appear adversarial while actually producing two
independently productive factions rather than a genuine arms race.

Our contributions are:
\begin{enumerate}
  \item \textbf{PGG arms race.} 30~generations of alternating
    co-evolution drive Blue cooperators from $S_B{=}0.370$ to $0.777$ and
    Red free-riders from $S_R{=}0.177$ to $0.782$, converging to a stable
    near-parity equilibrium robust across tested multipliers
    $m \in \{1.2, 1.5, 2.0, 3.0\}$ (Section~\ref{sec:exp4}).

  \item \textbf{Fitness coupling as a precondition for adversarial dynamics
    in independently-scored environments.} Per-faction scoring inherited
    from~\cite{kumar2026evolving} leaves outcomes statistically uncoupled
    when extended to two factions ($\text{corr}(S_B, S_R){=}{+}0.088$,
    $S_B{+}S_R$ varying). We propose score-advantage fitness
    $S_{\text{own}}{-}S_{\text{opp}}$ as the search target and show this
    single change enables measurable adversarial dynamics
    (Section~\ref{sec:protocol}).

  \item \textbf{Grid-world arms race under the corrected protocol.}
    $C^*$ retains its lead under adversarial search (mean Red advantage
    $-0.27$); full co-evolution converges near parity; asymmetric
    information lets Red dominate (final Red advantage $+0.415$); a
    coordination requirement on attacks drops Red's mean advantage to
    $-0.66$ without modifying the cooperator's constitution
    (Sections~\ref{sec:exp1}--\ref{sec:exp3}).

  \item \textbf{Evaluation seed count $K$ controls mode-regression in
    LLM-guided pure-adversary search.} At $K{=}2$ search regresses; at
    $K{=}5$ Red sustains a strong specialist for all 30~generations,
    isolating fitness-estimate noise as the readily-controllable lever
    (Section~\ref{sec:exp5}).
\end{enumerate}

We do not claim that adversarially-evolved cooperative constitutions are
more robust than cooperatively-evolved ones; testing that hypothesis would
require a transfer experiment we have not run. Instead, this paper
contributes a methodology for constructing and diagnosing constitutional arms
races, showing when adversarial co-evolution works, when it silently fails,
and what kinds of interpretable adversarial constitutions emerge under
cooperation--defection pressure. This reframes constitutional alignment not
only as a problem of designing good principles, but also as a problem of
stress-testing governance rules against adaptive social conflict.

\section{Background and Related Work}

\paragraph{Constitutional AI and multi-agent alignment.}
CAI aligns models using human-written principles and AI
feedback~\citep{bai2022constitutional, askell2021general}.
Extensions have improved helpfulness and reduced harmful outputs but maintain
the single-agent paradigm. \citet{carichon2025crisis} argue that alignment
must become a ``dynamic and social process'' for multi-agent settings.
\citet{kumar2026evolving} operationalize this through evolutionary search
over cooperative constitutions; we extend that line to adversarial
co-evolution.

\paragraph{Multi-agent coordination and self-play.}
\citet{axelrod1981evolution} showed that tit-for-tat evolves stable
cooperation in iterated games; \citet{leibo2017multi} introduced sequential
social dilemmas in multi-agent RL; \citet{hughes2018inequity}
and \citet{jaques2019social} demonstrated that social preferences and
influence rewards stabilize cooperation. \citet{park2023generative} showed
emergent social behaviors in LLM agents. Competitive self-play trains strong
policies in two-player games~\citep{silver2016mastering, openai2019dota5};
however, self-play optimizes policy parameters rather than a natural-language
governance artifact. \citet{wang2019poet} co-evolve agents and environments
in single-agent curricula; we adapt this co-evolutionary structure to
two-faction constitutional search. We do not update policy weights or
generate code: we evolve interpretable priority-ordered natural-language
rules that LLM agents follow verbatim at decision time.

\paragraph{Mechanism design and LLM-guided evolutionary search.}
Our framework relates to automated mechanism
design~\citep{sankar2024deep, liu2025interpretable, koster2022human}, where
rules governing agent interaction are optimized for desired outcomes; we
operate in partially observable environments where agent utility functions
are implicit in LLM behavior rather than explicitly specified. The search
machinery itself follows the FunSearch~\citep{romera2024mathematical} and
AlphaEvolve~\citep{alphaevolve2025} line of LLM-as-mutation-operator
program synthesis, with quality-diversity through
ELM~\citep{lehman2022elm} and OpenEvolve~\citep{openevolve2025} on
MAP-Elites~\citep{mouret2015mapelites}. A common assumption in this line is
that the LLM mutation operator faithfully implements the stated objective;
our mode-regression result (Section~\ref{sec:exp5}) documents a case where
this breaks under fitness-estimate noise.

\paragraph{Public goods games.}
The PGG is a canonical model for cooperation and free-riding.
\citet{fehr2000cooperation} showed that costly punishment sustains
cooperation; \citet{hauert2002volunteering} showed that voluntary
participation introduces cyclic dynamics; \citet{rand2009positive} showed
that positive interactions promote cooperation. We use the PGG as a testbed
for constitutional co-evolution that inherently provides the fitness
coupling absent from independently-scored grid-world experiments.

\section{Methodology}

\subsection{Constitutional Co-Evolution}

We consider a multi-agent setting where $n$ agents are divided into two
factions---Blue and Red. Each faction is governed by a constitution
$C = \{r_1, \ldots, r_k\}$, a priority-ordered set of natural-language rules
that its agents follow at decision time. Faction observability is
environment-dependent: in the grid-world, faction identities are hidden from
agents, while in the PGG, faction roles are induced by the initialized
cooperative or free-riding constitutions. Adversarial
co-evolution produces a sequence of constitution pairs $(C^g_B, C^g_R)$ for
$g = 1, \ldots, G$, where in each generation each faction's constitution is
updated to maximize its expected score against the most recent opponent
constitution. We approximate each maximization with one iteration of
LLM-guided evolutionary search via OpenEvolve~\citep{openevolve2025} on top
of MAP-Elites~\citep{mouret2015mapelites}, alternating between Blue and Red.
Three fitness targets appear in our experiments: \emph{absolute}
($S_{\text{faction}}$), \emph{advantage} ($S_{\text{faction}}{-}S_{\text{opp}}$),
and \emph{pure-adversary} ($1{-}S_{\text{opp}}$, Red only). The reported
stability score $S$ is identical in all modes; only the optimization target
changes. The full algorithm and formal optimization definition are given in
Appendix~\ref{app:algorithm}.

\subsection{Faction-Specific Scoring}

We adapt the cooperative scoring of~\citet{kumar2026evolving} to the
two-faction setting. Each faction is scored independently:
\begin{equation}
    S_\text{faction} = \alpha \cdot P + \beta \cdot V - \gamma \cdot C_\text{ff}
    \label{eq:score}
\end{equation}
where $P \in [0,1]$ is productivity (faction-specific project progress in
grid-world, pool utilization in PGG), $V \in [0,1]$ is survival rate, and
$C_\text{ff} \in [0,1]$ is the \emph{friendly-fire} rate. We use $\alpha{=}0.5$,
$\beta{=}0.3$, $\gamma{=}0.2$ throughout.

Two changes from the cooperative scoring of~\citet{kumar2026evolving} are
worth flagging:
(i)~the conflict term $C$ in the prior work counted all aggressive actions;
in the adversarial setting we restrict it to within-faction aggression
($C_\text{ff}$), since cross-faction attacks are a legitimate strategic tool;
(ii)~the prior work clipped $S$ at zero; we allow $S < 0$ so that the search
gradient is preserved when a faction is being dominated. Absolute $S$ values
in this paper are therefore not directly comparable to those
in~\cite{kumar2026evolving}; only relative trends and inter-faction gaps are.

\section{Environments}

\subsection{Grid-World with Hidden Adversarial Identity}
\label{sec:gridworld}

We adapt the grid-world primitives of~\citet{kumar2026evolving} to a
two-faction adversarial setting. The cooperative environment in that work
divided six agents into two cooperating teams sharing a single constitution
and a joint success criterion. We retain the resource types, action
primitives, and per-step dynamics, and modify the setup as follows: two
adversarial factions (Blue, Red) replace the cooperating teams, each with
its own constitution; faction labels are hidden from agents; Blue and Red
have symmetric project pairs (Shelter $+$ Market vs.\ Fortress $+$ Armory,
identical resource requirements: 150 wood, 120 stone, 30 gems each); the
grid is scaled from $6{\times}6$ to $8{\times}8$ to give two factions of
three agents each room to operate; episodes are 40~turns; the action set
(gather, deposit, move, communicate, attack at $p{=}0.25$, steal at
$p{=}0.40$, give, rest) is inherited from the prior work; and the periodic
Overseer-elimination and hunger mechanics are disabled, since they confound
adversarial scoring (Appendix~\ref{app:env-changes} details the rationale).
We do not claim absolute comparability of $S$ values to~\cite{kumar2026evolving};
the grid-world here is a related but distinct testbed.

\subsection{Public Goods Game}
\label{sec:pgg_env}

To evaluate the method in a non-spatial social dilemma, we introduce
a structurally different environment based on the classic
PGG~\citep{fehr2000cooperation}. Six agents (3 Blue cooperators, 3 Red
free-riders) play 20~rounds with multiplier $m{=}1.5$ and initial endowment
100 per agent. Each round, agents choose to contribute (0--100\% of wallet),
punish (cost ratio $0.3$ to punisher, $1.0$ impact on target), broadcast a
message, or rest. Scoring uses the same Eq.~\eqref{eq:score} formula with
$P =$ pool utilization, $V = 1.0$ (no death in PGG), and $C_\text{ff} =$
within-faction free-riding rate.

The PGG captures a fundamentally different social dilemma: no spatial
structure, no combat, no resource gathering. The shared pool inherently
couples both factions' payoffs --- contributions feed the same pot, which
is multiplied by $m$ and split equally --- creating zero-sum strategic
pressure without requiring any modification of the per-faction scoring.

\section{Experiments}

\textbf{Organization.} We present the main PGG result first
(Section~\ref{sec:exp4}), then the PGG ablations
(Sections~\ref{sec:pgg-pure-adv}, \ref{sec:pgg-mult}). We then turn to the
grid-world. Adversarial dynamics in the grid-world require a fitness-coupling
correction (Section~\ref{sec:protocol}), which we develop empirically before
reporting Experiments~1--3 (Sections~\ref{sec:exp1}--\ref{sec:exp3}) and a
defensive variant (Section~\ref{sec:exp1-coord}). A final ablation on
evaluation seed count (Section~\ref{sec:exp5}) isolates fitness-noise as the
main driver of mode-regression in pure-adversary search.

\subsection{PGG Co-Evolution}
\label{sec:exp4}

We run 30~generations of alternating co-evolution on the standard PGG
($N{=}6$, $m{=}1.5$, 20~rounds, endowment 100, $K{=}2$ seeds) with Blue
initialized from a cooperator seed (generous contribution, punish
free-riders) and Red from a free-rider seed (minimal contribution, deceptive
signaling). Seed constitutions are in Appendix~\ref{app:constitutions}.

\begin{figure}[t]
    \centering
    \includegraphics[width=\textwidth]{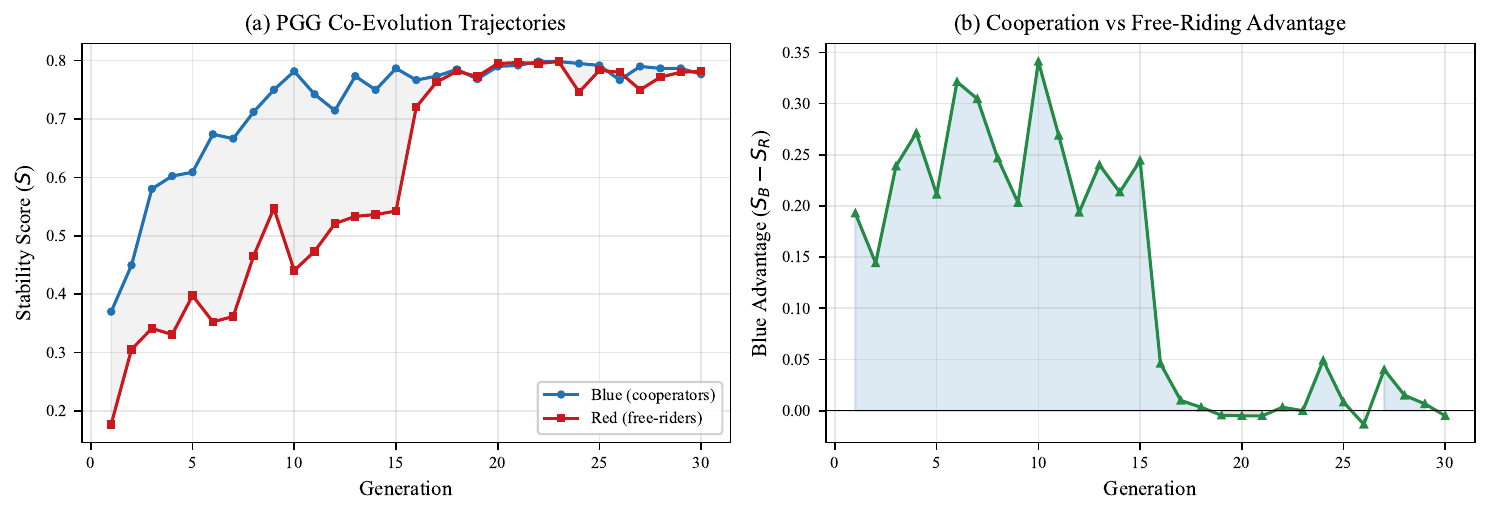}
    \caption{PGG co-evolution over 30 generations. \textbf{(a)}~Blue
    (cooperators) and Red (free-riders) stability scores; both improve and
    converge near $S \approx 0.78$. \textbf{(b)}~Blue advantage
    ($S_B - S_R$): large early lead collapses toward zero as Red's
    constitution learns to survive punishment, settling at near-parity.}
    \label{fig:pgg}
\end{figure}

Both factions improve substantially. Blue climbs from $S_B{=}0.370$ to
$S_B{=}0.777$ ($+110\%$) and Red from $S_R{=}0.177$ to $S_R{=}0.782$
($+341\%$). Red catches up to Blue by generation~18, and both converge at a
stable near-parity equilibrium around $S \approx 0.78$ for the final ten
generations (Table~\ref{tab:pgg-phases}, Figure~\ref{fig:pgg}).

\begin{table}[h]
\centering
\caption{PGG co-evolution trajectory: 5-generation window endpoints. The
score gap collapses to near zero by generation~20 and stays stable through
generation~30.}
\label{tab:pgg-phases}
\begin{tabular}{llll}
\toprule
\textbf{Phase} & \textbf{Gens} & \textbf{Blue $S$} & \textbf{Red $S$} \\
\midrule
Early climb  & 1--5   & $0.370 \rightarrow 0.609$ & $0.177 \rightarrow 0.398$ \\
Blue leads   & 6--10  & $0.674 \rightarrow 0.782$ & $0.353 \rightarrow 0.441$ \\
Red catch-up & 11--15 & $0.742 \rightarrow 0.792$ & $0.473 \rightarrow 0.541$ \\
Convergence  & 16--20 & $0.770 \rightarrow 0.792$ & $0.720 \rightarrow 0.797$ \\
Plateau      & 21--30 & $0.767 \rightarrow 0.777$ & $0.750 \rightarrow 0.782$ \\
\bottomrule
\end{tabular}
\end{table}

The PGG pool is inherently shared: all agents' contributions feed the same
pot, which is multiplied by $m$ and split equally. This structural coupling
creates adversarial selection pressure without requiring any modification of
the per-faction fitness function --- Red must discover that some level of
contribution is necessary to avoid Blue's punishment, and Blue must discover
that punishment is necessary to sustain high contribution. The convergence
at $S \approx 0.78$ is consistent with the high-cooperation rates reported
by~\citet{fehr2000cooperation} for PGGs with costly punishment, though we
note that $S$ aggregates productivity, survival, and friendly-fire into a
single scalar and is not a direct measure of contribution rate. The
headline trajectory is a single $K{=}2$ run; the multiplier ablation
(Section~\ref{sec:pgg-mult}) provides evidence that the equilibrium target
is structural rather than seed-specific.

\subsection{PGG under Pure-Adversary Fitness}
\label{sec:pgg-pure-adv}

A natural question is whether the convergence above depends on Red having
its own productivity to optimize. We rerun the PGG with Red's fitness target
set to $1{-}S_B$ (pure-adversary, no productivity reward) starting from a
pure-adversary seed. Across the 21~valid generations of a 30-generation run
(gens~2--10 returned $S_B{=}S_R{=}0$ from a recoverable LLM-mutation
syntax-error pathology, Appendix~\ref{app:preliminary}), Red maintains tight
Blue suppression: $S_B$ stays in $[0.174, 0.286]$ with mean $0.198$, and
Red fitness averages $0.802$, with final fitness ($0.798$) essentially equal
to gen-1 fitness ($0.822$). This contrasts with the grid-world (Section~\ref{sec:exp5}),
where the same fitness mode at $K{=}2$ exhibits mode-regression. We attribute
the difference to PGG's small action space (contribute, punish, broadcast,
rest) compared to grid-world's wider behavioral surface; with fewer
non-adversarial actions available, the mutation operator's productive-action
prior has less material to drift toward. \emph{Environment structure}, not
just fitness signal, shapes whether adversarial search remains stable. The
full per-generation trajectory is in Appendix~\ref{app:pgg-pure-adv}.

\subsection{Multiplier Ablation: $m \in \{1.2, 1.5, 2.0, 3.0\}$}
\label{sec:pgg-mult}

To test whether the near-parity equilibrium is specific to $m{=}1.5$, we
rerun the same setup at $m \in \{1.2, 2.0, 3.0\}$, spanning the
social-dilemma range ($N{=}6$, so $m{=}1.2$ sits near the lower bound and
$m{=}3.0$ near the upper). All four runs converge to near-parity with mean
Red advantage in $[-0.027, +0.005]$ over the final five generations
(Table~\ref{tab:pgg-mult}). The multiplier affects only the transient: at
$m{=}1.2$ both factions start low ($S_B{=}0.341$, $S_R{=}0.170$ at gen~1)
and climb together; at $m{=}3.0$ Red's free-rider seed is heavily punished
early ($S_R{=}0.287$ at gen~1) and catches up sharply by gen~5
($S_R{=}0.727$). The steady state is the same high-cooperation equilibrium
across all four conditions, indicating the result reflects the structural
property of pool-based payoff coupling rather than a knife-edge multiplier
calibration.

\begin{table}[h]
\centering
\caption{PGG multiplier ablation: final-generation stability scores. All
four multipliers converge to $S \approx 0.78$.
$^*$The $m{=}1.2$ run completed 26 of 30 generations at submission, stabilized
in $[0.78, 0.80]$ for the last 10.}
\label{tab:pgg-mult}
\begin{tabular}{lcccc}
\toprule
\textbf{$m$} & \textbf{Gens} & \textbf{Final $S_B$} & \textbf{Final $S_R$} & \textbf{Mean Red adv. (last 5)} \\
\midrule
1.2 (near PD)   & 26/30$^*$ & 0.782 & 0.797 & $+0.005$ \\
1.5 (main)      & 30/30     & 0.777 & 0.782 & $+0.005$ \\
2.0             & 30/30     & 0.785 & 0.767 & $-0.018$ \\
3.0 (near coop) & 30/30     & 0.778 & 0.800 & $-0.027$ \\
\bottomrule
\end{tabular}
\end{table}

\subsection{Fitness-Coupling Correction for Grid-World Experiments}
\label{sec:protocol}

The per-faction scoring in Eq.~\eqref{eq:score} references only within-faction
quantities, so optimizing $S_R$ independently yields no gradient toward
strategies that degrade the opponent. We confirm this empirically: with Blue
fixed at $C^*$ from~\cite{kumar2026evolving} and Red evolving from a
zero-sum seed under independent scoring, Pearson correlation between $S_B$
and $S_R$ across 30~generations is $+0.088$, and the sum $S_B{+}S_R$ varies
over $[1.09, 1.31]$ --- both inconsistent with a zero-sum game. Red's $S_R$
rises from $0.483$ to $0.602$ at generation~7 and plateaus in $[0.49, 0.60]$
thereafter, indicating Red has evolved productive self-building behavior
rather than adversarial behavior (full diagnostic in
Appendix~\ref{app:preliminary}).

We therefore adopt a \emph{score-advantage} fitness target for the
grid-world experiments:
\begin{equation}
    \text{fitness}_{\text{faction}} = S_{\text{faction}} - S_{\text{opponent}}.
    \label{eq:advantage}
\end{equation}
The reported stability score $S$ (Eq.~\eqref{eq:score}) is unchanged; only
the search target changes. The formulation is symmetric, so the
cooperative-vs-adversarial asymmetry is carried entirely by the initial
seed constitutions.

\subsection{Grid-World Experiment 1: $C^*$ Fragility under Adversarial Pressure}
\label{sec:exp1}

We fix Blue at $C^*$, the best cooperative constitution
from~\cite{kumar2026evolving}, and let Red evolve from a zero-sum seed for
30~generations under score-advantage fitness ($8{\times}8$ grid, 40~turns,
$K{=}2$ seeds, hidden faction identity).

$C^*$ exhibits structural resilience throughout. Red's advantage
$S_R - S_B$ remains negative for all 30~generations; the mean is $-0.27$
and the best (least negative) is $-0.079$ at generation~10
(per-generation values and trajectory plot in
Appendix~\ref{app:gridworld-tables}). $C^*$'s operational specificity
(rules like ``deposit immediately if carrying a needed resource'') confers
adversarial robustness at this compute budget: Red closes the gap to within
$0.08$ but does not surpass it, and the trajectory shows no monotonic
degradation. We do not claim that $C^*$ is robust in general --- only that
30~generations of LLM-guided search at $K{=}2$ does not break it. Stronger
adversarial seeds, larger $K$, or longer runs may yield different
conclusions.

\subsection{Defensive Variant: Coordination-Required Attacks}
\label{sec:exp1-coord}

A natural defensive question: does the cooperative equilibrium become harder
to break if attacks themselves require coordination? We rerun Exp~1 with
attack success gated on faction adjacency: solo attacks succeed at
$p{=}0.05$ (vs.\ $0.25$ baseline), coordinated attacks (one or more same-faction
allies adjacent to the target) succeed at $p{=}0.60$. The change is
symmetric but matters only to Red, since $C^*$ does not attack. Across
30~generations, Red's mean advantage drops to $-0.663$ (vs.\ $-0.27$ in
the baseline Exp~1), and Red advantage is $\geq 0$ in only 1 of
30~generations. Red's evolutionary search does discover that clustering
attackers raises success, but coordinating multiple agents on the same
target simultaneously across an $8{\times}8$ grid with hidden faction
identity is a coordination problem the LLM-mutation operator does not
reliably solve. This points to a practical lever: \emph{spatial or temporal
coordination requirements on damaging actions} raise adversarial search
difficulty without changing the fitness landscape or the cooperator's
rules. Full per-generation values in Appendix~\ref{app:gridworld-tables}.

\subsection{Grid-World Experiment 2: Full Co-Evolution}
\label{sec:exp2}

We now let both factions evolve under symmetric score-advantage fitness:
Blue from $C^*$ seed, Red from zero-sum seed, both optimizing
$S_{\text{own}} - S_{\text{opp}}$ in alternating mode (30~generations,
$K{=}2$). The trajectory unfolds in three phases. In an early phase
(gens~1--12), Blue holds a lead with Red advantage averaging $-0.18$;
through gens~13--20 the gap closes; by gens~21--30 both factions
oscillate near parity, with Red briefly establishing a lead at gens~26--27
before Blue partially recovers. Final values: $S_B{=}0.191$, Red advantage
$+0.192$. The mean Red advantage across all 30~generations is $-0.005$,
consistent with zero-sum search at near-parity. Score-advantage fitness
produces a competitive equilibrium rather than the paralysis observed
under independent scoring (where both factions collapsed toward zero
within 15~generations in pilot runs). The absolute $S$ values are low
because $C_\text{ff}$ and the absence of the $\max(0,\cdot)$ floor allow
negative scores; what matters here is the inter-faction gap, which
oscillates around zero. Per-generation values and the four-panel
trajectory plot in Appendix~\ref{app:gridworld-tables}.

\subsection{Grid-World Experiment 3: Information Asymmetry Ablation}
\label{sec:exp3}

We rerun Exp~2 under two information conditions:
(i)~\textbf{full}---both factions observe all six agents' action logs;
(ii)~\textbf{asymmetric}---Red observes all six agents, Blue observes only
its own three. A third \textbf{own-team} condition (each faction sees only
its own agents) hit the LLM-mutation syntax-error pathology before
completion (Appendix~\ref{app:preliminary}) and is not reported.

\begin{figure}[h]
    \centering
    \includegraphics[width=0.85\textwidth]{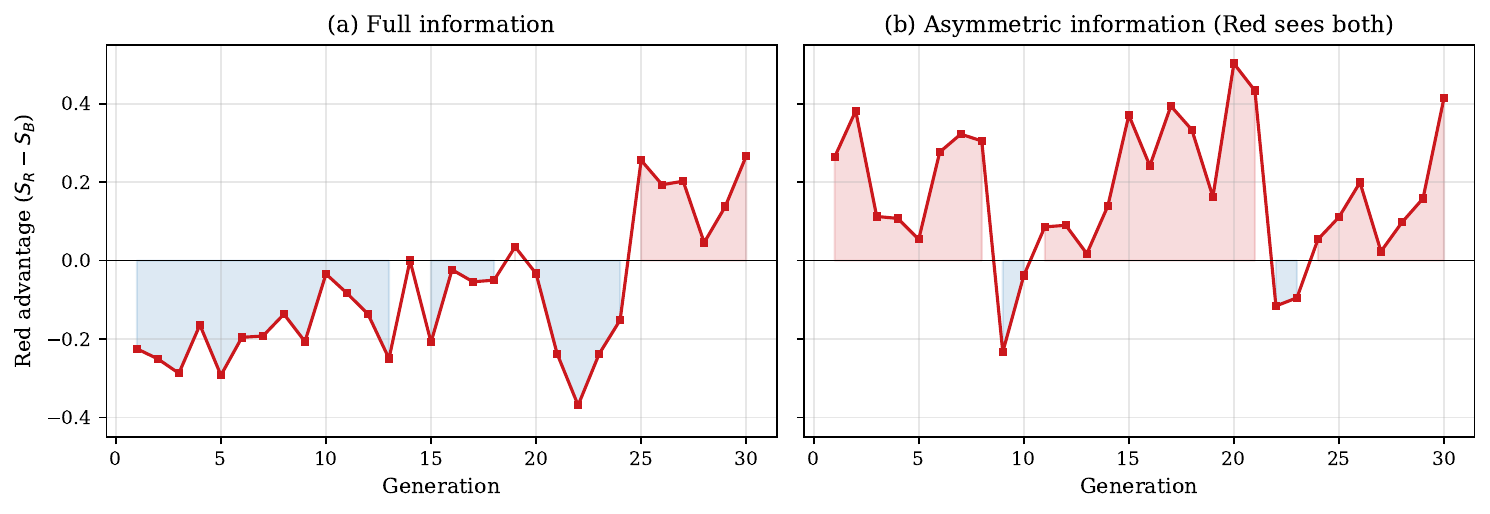}
    \caption{Information asymmetry (Experiment~3). Red advantage
    $S_R{-}S_B$ over 30~generations of full co-evolution.
    \textbf{(a)}~Full information: Blue leads through generation~24, then
    Red establishes a sustained advantage.
    \textbf{(b)}~Asymmetric information (Red observes both factions, Blue
    observes own team only): Red leads from generation~1 and dominates in
    25~of 30~generations, peaking at $+0.503$.}
    \label{fig:info_asymmetry}
\end{figure}

Information availability substantially shifts the equilibrium
(Figure~\ref{fig:info_asymmetry}). Under \textbf{full information}, Blue
holds an early lead (Red advantage averaging $-0.18$ through gen~24) before
Red establishes a sustained lead in the final six generations (final Red
advantage $+0.267$). Under \textbf{asymmetric information}, Red leads
immediately ($+0.264$ at gen~1), reaches a maximum advantage of $+0.503$ at
gen~20, drives Blue to $S_B{=}-0.367$ at gen~17, and ends with Red advantage
$+0.415$. With sustained intel disadvantage, Blue evolves rules that fail
against the specific Red strategies it cannot fully observe. The final
$+0.415$ in the asymmetric condition is the strongest Red result we observe
in any grid-world experiment. Per-generation values for both conditions in
Appendix~\ref{app:gridworld-tables}.

\subsection{Evaluation Seed Count and Mode-Regression in Pure-Adversary Search}
\label{sec:exp5}

When Red has no productive goals of its own --- fitness target $1 - S_B$,
seed constitution restricted to adversarial rules --- does evolutionary
search converge on a stable specialist? With Blue fixed at $C^*$ and Red
evolving from a 5-rule pure-adversary seed (hunt Blue, steal to disrupt,
identify teammates, avoid friendly fire, no productive actions) for
30~generations, the answer at $K{=}2$ is no. As reference, a hand-coded
20-line hunt-and-kill controller (move to nearest non-ally, attack when
adjacent, no gathering) eliminates all Blue agents by turn~8 and yields
$S_B{=}0.000$, fitness $= 1.0$ --- an upper bound on achievable
pure-adversary performance.

At $K{=}2$, Red's fitness is non-monotonic: it peaks at generation~5
($S_B{=}0.129$, fitness $0.871$), then degrades to fitness $0.508$ by
generation~30 ($S_B{=}0.492$). From generation~15 onward, $S_R$ drifts
upward and rule count grows from 6 to 8--9 (Table~\ref{tab:exp5-trajectory}),
indicating re-acquisition of productive rules despite a seed that prohibits
them and a fitness signal that does not reward them. Across all
30~generations, Red never approaches the adversarial capability of the
hand-coded controller. This gap is real, and we do not paper over it:
LLM-guided constitutional search underperforms a trivial specialist on raw
adversarial capability. The value our framework adds lies elsewhere --- the
evolved Red constitutions are interpretable natural-language rules that can
be inspected, transferred across seeds, and applied as red-team test cases
against unseen cooperative constitutions, none of which the hand-coded
controller offers.

\begin{table}[h]
\centering
\caption{Pure-adversary search: per-generation values binned into
5-generation windows. Fitness target is $1{-}S_B$; $S_R$ is reported but
not optimized. Hand-coded baseline: $S_B{=}0.000$, fitness $= 1.000$.}
\label{tab:exp5-trajectory}
\begin{tabular}{lcccc}
\toprule
\textbf{Window} & \textbf{mean $S_B$} & \textbf{mean fitness} & \textbf{mean $S_R$} & \textbf{mean rule count} \\
\midrule
Gen 1--5   & 0.284 & 0.716 & 0.290 & 5.6 \\
Gen 6--10  & 0.266 & 0.734 & 0.300 & 6.0 \\
Gen 11--15 & 0.288 & 0.712 & 0.311 & 5.8 \\
Gen 16--20 & 0.387 & 0.613 & 0.370 & 6.0 \\
Gen 21--25 & 0.352 & 0.648 & 0.386 & 7.0 \\
Gen 26--30 & 0.492 & 0.508 & 0.419 & 8.6 \\
\bottomrule
\end{tabular}
\end{table}

Two ablations isolate the cause. \textbf{(1)~$K{=}5$, 30~generations}:
increasing the evaluation seed count to $K{=}5$ eliminates mode-regression
entirely. Red sustains adversarial pressure throughout: $S_B$ remains in
$[0.13, 0.40]$ for all 30~generations, with peak $S_B{=}0.127$ at gen~18
and final fitness $0.825$ --- substantially higher than the $K{=}2$
endpoint ($0.508$) and comparable to the $K{=}2$ peak ($0.871$).
\textbf{(2)~$G{=}60$, $K{=}2$}: extending the run to 60~generations shows
that mode-regression is delayed but not eliminated at small $K$. Red
reaches a deeper minimum at gen~23 ($S_B{=}0.083$, fitness $0.917$) before
drifting back; by gen~60, fitness returns to $0.516$.

The $K{=}5$ result indicates that fitness-estimate noise is a measurable
contributor to mode-regression: with five seeds the variance drops by
${\sim}2.5{\times}$, and regression no longer occurs. The $G{=}60$ result
indicates that the mutation operator's prior is also active --- even at
$K{=}2$, the search reaches its deepest specialist at gen~23 rather than
collapsing earlier. Mode-regression in LLM-guided evolution is best
understood as the joint effect of fitness noise and operator prior. For
adversarial-specialist search with LLM mutation, $K \geq 5$ should be
considered a default rather than $K{=}2$.

\section{Cross-Environment Analysis}

Across both environments under appropriately coupled fitness, we observe
common properties. Both factions improve over generations and arms race
dynamics emerge. Co-evolution discovers strategies that one-shot LLM design
does not: communication minimization in the grid-world,
punishment-calibrated contribution in the PGG. All evolved constitutions
remain human-readable priority-ordered rules. Environment structure
controls fitness-mode stability --- pure-adversary fitness regresses in the
grid-world but remains stable in the PGG, due to the narrower PGG action
space. Mechanism design functions as a defensive lever: a single
coordination requirement on attacks drops Red's mean advantage from $-0.27$
to $-0.66$ without modifying any constitution. Final stability scores
differ in absolute terms across environments (PGG: $S_B{=}0.777$,
$S_R{=}0.782$; grid-world Exp~2: $S_B{=}0.191$, $S_R{=}0.192$;
Exp~3-asymmetric: final Red advantage $+0.415$), but absolute $S$
comparisons across environments are not meaningful because $S$ aggregates
different productivity components in each setting.

\section{Discussion}

\paragraph{Fitness coupling is a precondition for adversarial dynamics.}
A central methodological contribution of this work is identifying that the
natural per-faction scoring $S_\text{faction}$ does not couple the two
factions' outcomes in independently-scored environments. Optimizing $S_R$
independently produces a productive second faction rather than an
adversary. The PGG's shared-pool structure provides this coupling natively
--- and the clean arms race dynamics in Section~\ref{sec:exp4}, robust
across $m \in \{1.2, 1.5, 2.0, 3.0\}$, suggest that environments with
explicit resource sharing may serve as more reliable testbeds for adversarial
constitutional co-evolution than independently-scored games. For grid-world
environments the coupling must be imposed explicitly via score-advantage
fitness (Eq.~\eqref{eq:advantage}). This is an instance of a broader
principle in mechanism design: the game's payoff structure determines what
evolutionary pressure actually selects for.

\paragraph{Evaluation seed count as a practical lever.}
Section~\ref{sec:exp5} surfaces evaluation seed count $K$ as a key
hyperparameter for LLM-guided adversarial search: increasing $K$ from 2
to~5 eliminates mode-regression by reducing fitness-estimate noise. In
contrast to prior work focused on mutation operator
design~\citep{romera2024mathematical, alphaevolve2025}, this points to
evaluation budget as an equally important axis.

\paragraph{Evolved Red constitutions as red-team artifacts.}
The adversarial Red constitutions are themselves useful artifacts: each is
a priority-ordered set of natural-language rules that produces measurable
damage against a known cooperative constitution. The
asymmetric-information Red (Exp~3, final advantage $+0.415$) and the
$K{=}5$ pure-adversary Red (Section~\ref{sec:exp5}, sustained
$S_B \in [0.13, 0.40]$) are concrete red-team test cases that can be
applied to evaluate future constitutional designs. Unlike adversarial
neural policies, these artifacts are interpretable: practitioners can
inspect the rules an attacker would follow and reason about which
constitutional clauses they exploit.

\paragraph{What we do not claim.}
We do not claim that adversarially-evolved cooperative constitutions are
more robust than cooperatively-evolved ones. Testing that hypothesis
would require running an adversarially-evolved Blue against a fresh
adversarial Red and comparing against $C^*$ --- an experiment we have not
performed.

\paragraph{Limitations and broader impact.}
Our framework relates to automated mechanism
design~\citep{sankar2024deep, liu2025interpretable} where the ``mechanism''
is a natural-language constitution; the coordination-required attack result
(Section~\ref{sec:exp1-coord}) shows that mechanism and constitutional
design are complementary. All headline trajectories are single $K{=}2$
runs; variance across independent runs is not characterized. Two
conditions hit a recoverable LLM-mutation syntax-error pathology and are
reported with caveats; reruns are planned. The grid-world here differs
from~\cite{kumar2026evolving} in scoring and several mechanics
(Appendix~\ref{app:env-changes}), so absolute $S$ values are not directly
comparable. Discovering governance rules through evolutionary optimization
requires ongoing human oversight, and the adversarial Red constitutions
are interpretable attack artifacts whose research value as red-team test
cases must be weighed against misuse risk before any operational
deployment.

\section{Conclusion}

Adversarial constitutional co-evolution works in environments with native
fitness coupling: in the PGG, 30~generations drive both factions to a
near-parity equilibrium at $S \approx 0.78$, robust across multipliers.
In independently-scored environments the natural setup produces no
adversarial pressure ($\text{corr}(S_B, S_R){=}{+}0.088$); a
score-advantage fitness target restores it. The same operator regresses
at $K{=}2$ and stabilizes at $K{=}5$, identifying evaluation noise as the
primary controllable lever. The evolved Red constitutions are themselves
interpretable red-team artifacts for future cooperative constitutions.

\bibliography{references}


\newpage
\appendix

\section*{Appendix}
\addcontentsline{toc}{section}{Appendix}

\section{Formal Algorithm and Optimization Definition}
\label{app:algorithm}

The framework summary in Section~\ref{sec:exp4} omitted formal details for
brevity. Here we give them explicitly.

\begin{definition}[Co-Evolution]
Given environment $\mathcal{E}$, initial constitutions $C^0_B, C^0_R$, and
scoring function $S$, adversarial co-evolution produces a sequence of
constitution pairs $(C^g_B, C^g_R)$ for $g = 1, \ldots, G$ where each
generation $g$:
\begin{align}
    C^g_B &= \argmax_{C} \; \mathbb{E}_{\tau \sim p(\tau | C, C^{g-1}_R, \mathcal{E})}[S_B(\tau)] \\
    C^g_R &= \argmax_{C} \; \mathbb{E}_{\tau \sim p(\tau | C^g_B, C, \mathcal{E})}[S_R(\tau)]
\end{align}
\end{definition}

We approximate each maximization with one iteration of LLM-guided
evolutionary search via OpenEvolve~\citep{openevolve2025}, alternating
between Blue and Red. Algorithm~\ref{alg:coev} gives the procedure.

\begin{algorithm}[h]
\caption{Adversarial Constitutional Co-Evolution (alternating mode)}
\label{alg:coev}
\begin{algorithmic}[1]
\Require Initial constitutions $C^0_B, C^0_R$; environment $\mathcal{E}$;
         generations $G$; evaluation seeds $K$;
         fitness mode $\phi \in \{\text{absolute}, \text{advantage}, \text{pure-adv}\}$
\Ensure Generation log $\mathcal{L} = \{(C^g_B, C^g_R, S^g_B, S^g_R)\}_{g=1}^{G}$
\For{$g = 1$ to $G$}
    \State \textbf{Evolve Blue:} $C^g_B \leftarrow$ \textsc{OpenEvolveStep}$\big(C^{g-1}_B,\ \text{opp}{=}C^{g-1}_R,\ \phi_B\big)$
    \State \textbf{Evolve Red:}  $C^g_R \leftarrow$ \textsc{OpenEvolveStep}$\big(C^{g-1}_R,\ \text{opp}{=}C^{g}_B,\ \phi_R\big)$
    \State Sample $K$ trajectories: $\tau_1,\ldots,\tau_K \sim p(\cdot | \mathcal{E}, C^g_B, C^g_R)$
    \State $\hat S^g_B \leftarrow \tfrac{1}{K}\sum_i S_B(\tau_i)$;\quad
           $\hat S^g_R \leftarrow \tfrac{1}{K}\sum_i S_R(\tau_i)$
    \State Append $(C^g_B, C^g_R, \hat S^g_B, \hat S^g_R)$ to $\mathcal{L}$
\EndFor
\State \textbf{return} $\mathcal{L}$
\end{algorithmic}
\end{algorithm}

\textsc{OpenEvolveStep} runs one iteration of MAP-Elites evolutionary
search~\citep{mouret2015mapelites} with an LLM mutation operator. The
fitness target $\phi$ determines the scalar OpenEvolve maximizes:
$\phi{=}\text{absolute}$ uses $S_{\text{faction}}$;
$\phi{=}\text{advantage}$ uses $S_{\text{faction}}{-}S_{\text{opp}}$;
$\phi{=}\text{pure-adv}$ (Red only) uses $1{-}S_{\text{opp}}$. The reported
stability score $S$ is the same in all modes; only the search target
changes.

\section{Grid-World Environment Changes from~\cite{kumar2026evolving}}
\label{app:env-changes}

The grid-world environment in this paper inherits primitives from the
cooperative grid-world of~\cite{kumar2026evolving} but differs in several
respects worth detailing for reproducibility.

\textbf{Adversarial factions.} The prior work divided six agents into two
cooperating teams (Shelter, Market) sharing a single constitution. We
replace this with two adversarial factions (Blue, Red), each with its own
constitution and project pair. Symmetric resource requirements ensure
neither faction has a structural advantage.

\textbf{Hidden faction identity.} Agents cannot see other agents' faction
labels. The prior work had no such restriction since all agents shared a
team-cooperation goal.

\textbf{Grid scale.} We scale from $6{\times}6$ to $8{\times}8$ to give
two factions of three agents each room to operate without forced contact
on every turn. With $6{\times}6$ and adversarial dynamics, encounters
become trivially frequent and the strategic surface compresses.

\textbf{Conflict scoring.} The prior work's $C$ counted all aggressive
actions. In our adversarial setting cross-faction attacks are a legitimate
strategic tool, so we restrict the conflict term to $C_\text{ff}$
(within-faction aggression only). Same-faction friendly fire remains
penalized.

\textbf{$\max(0,\cdot)$ floor removed.} The prior work clipped $S$ at
zero. We allow $S < 0$ so that the search gradient is preserved when a
faction is being dominated; this is why $S_B$ goes negative in some
experiments (e.g., $S_B{=}-0.367$ at gen~17 of Exp~3-asymmetric).

\textbf{Overseer disabled.} The periodic-elimination mechanic confounds
adversarial scoring --- a Red ``win'' from killing Blue agents is
indistinguishable from a Red ``win'' from Overseer elimination of Blue
agents. We disable it.

\textbf{Hunger disabled.} The optional starvation mechanic is removed to
isolate strategic dynamics from resource-based attrition.

\section{PGG Pure-Adversary: Full Trajectory}
\label{app:pgg-pure-adv}

This appendix provides the full per-generation trajectory referenced in
Section~\ref{sec:pgg-pure-adv}. Setup: same PGG configuration as
Section~\ref{sec:exp4} ($m{=}1.5$, 30~generations, $K{=}2$ seeds), but
Red is initialized from a pure-adversary seed (5~rules: never contribute,
punish only Blue cooperators, avoid punishing free-riders, broadcast
targets to coordinate, treat the public pool as the opponent's resource).
Red's fitness target is $1{-}S_B$.

Across 30 generations, gens~2--10 returned $S_B{=}S_R{=}0$ from the
LLM-mutation syntax-error pathology described in
Appendix~\ref{app:preliminary}. We report selected valid generations in
Table~\ref{tab:pgg-pure-adv}.

\begin{table}[h]
\centering
\caption{PGG under pure-adversary Red fitness ($\text{fitness}_R{=}1{-}S_B$).
Selected valid generations from a 30-generation run; gens~2--10 are
excluded due to a recoverable syntax-error pathology. Red sustains Blue
suppression at $S_B \approx 0.20$ throughout.}
\label{tab:pgg-pure-adv}
\begin{tabular}{cccc}
\toprule
\textbf{Generation} & \textbf{$S_B$} & \textbf{$S_R$} & \textbf{Red fitness $1{-}S_B$} \\
\midrule
1  & 0.178 & 0.823 & 0.822 \\
15 & 0.201 & 0.805 & 0.799 \\
20 & 0.194 & 0.816 & 0.806 \\
25 & 0.211 & 0.784 & 0.789 \\
30 & 0.202 & 0.794 & 0.798 \\
\bottomrule
\end{tabular}
\end{table}

\section{Grid-World Per-Generation Trajectories}
\label{app:gridworld-tables}

This appendix collects the per-generation tables omitted from
Sections~\ref{sec:exp1}--\ref{sec:exp3} for space. All runs use $K{=}2$
evaluation seeds.

\begin{table}[h]
\centering
\caption{Exp~1 ($C^*$ fragility under score-advantage fitness). Blue is
fixed at $C^*$; Red evolves. Red never achieves a positive advantage.
Mean Red advantage $-0.27$; max (least negative) $-0.079$ at gen~10.}
\label{tab:exp1-trajectory}
\begin{tabular}{ccc}
\toprule
\textbf{Generation} & \textbf{$S_B$} & \textbf{Red advantage $S_R{-}S_B$} \\
\midrule
1  & 0.725 & $-0.321$ \\
5  & 0.642 & $-0.267$ \\
10 & 0.537 & $-0.079$ \\
15 & 0.717 & $-0.312$ \\
20 & 0.558 & $-0.246$ \\
25 & 0.714 & $-0.318$ \\
30 & 0.566 & $-0.355$ \\
\bottomrule
\end{tabular}
\end{table}

\begin{figure}[h]
    \centering
    \includegraphics[width=0.85\textwidth]{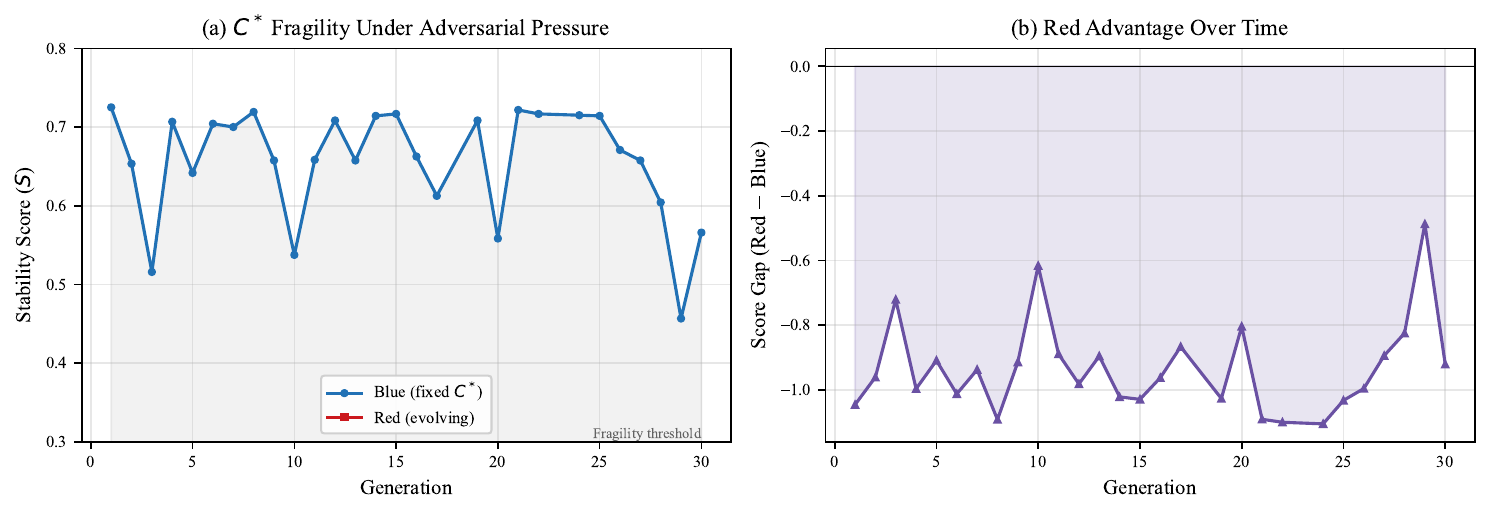}
    \caption{$C^*$ fragility (Experiment~1). \textbf{(a)}~Raw stability
    scores: Blue (fixed $C^*$) holds steadily across 30~generations of Red
    adversarial evolution. \textbf{(b)}~Red advantage $S_R - S_B$ remains
    negative throughout; $C^*$ retains its lead under sustained zero-sum
    adversarial search at this compute budget.}
    \label{fig:fragility}
\end{figure}

\begin{table}[h]
\centering
\caption{Coordination-required attacks (defensive variant of Exp~1). Solo
attack success $p{=}0.05$; coordinated success $p{=}0.60$. Red's advantage
is negative in 29 of 30 generations, with mean $-0.663$ (std 0.270).
Maximum Red advantage $0.000$ (gen~13); maximum Red-S $+0.122$ (gen~10).}
\label{tab:exp1-coord}
\begin{tabular}{ccc}
\toprule
\textbf{Generation} & \textbf{$S_B$} & \textbf{Red advantage $S_R{-}S_B$} \\
\midrule
1  & 0.707 & $-0.969$ \\
5  & 0.628 & $-0.937$ \\
10 & 0.382 & $-0.261$ \\
15 & 0.620 & $-0.716$ \\
20 & 0.535 & $-0.530$ \\
25 & 0.556 & $-0.607$ \\
30 & 0.479 & $-0.400$ \\
\bottomrule
\end{tabular}
\end{table}

\begin{table}[h]
\centering
\caption{Exp~2 (full co-evolution under score-advantage fitness): selected
generations. Three-phase arms race with initial Blue lead, progressive
equalization, and near-parity oscillation. Mean Red advantage across all
30~generations: $-0.005$.}
\label{tab:exp2-trajectory}
\begin{tabular}{ccc}
\toprule
\textbf{Phase / Gen} & \textbf{$S_B$} & \textbf{Red advantage} \\
\midrule
Gen 1   & $+0.344$ & $-0.192$ \\
Gen 6   & $-0.054$ & $+0.321$ \\
Gen 12  & $+0.043$ & $-0.106$ \\
Gen 18  & $-0.035$ & $+0.025$ \\
Gen 22  & $+0.268$ & $+0.253$ \\
Gen 26  & $-0.217$ & $+0.194$ \\
Gen 30  & $+0.191$ & $+0.192$ \\
\bottomrule
\end{tabular}
\end{table}

\begin{figure}[h]
    \centering
    \includegraphics[width=\textwidth]{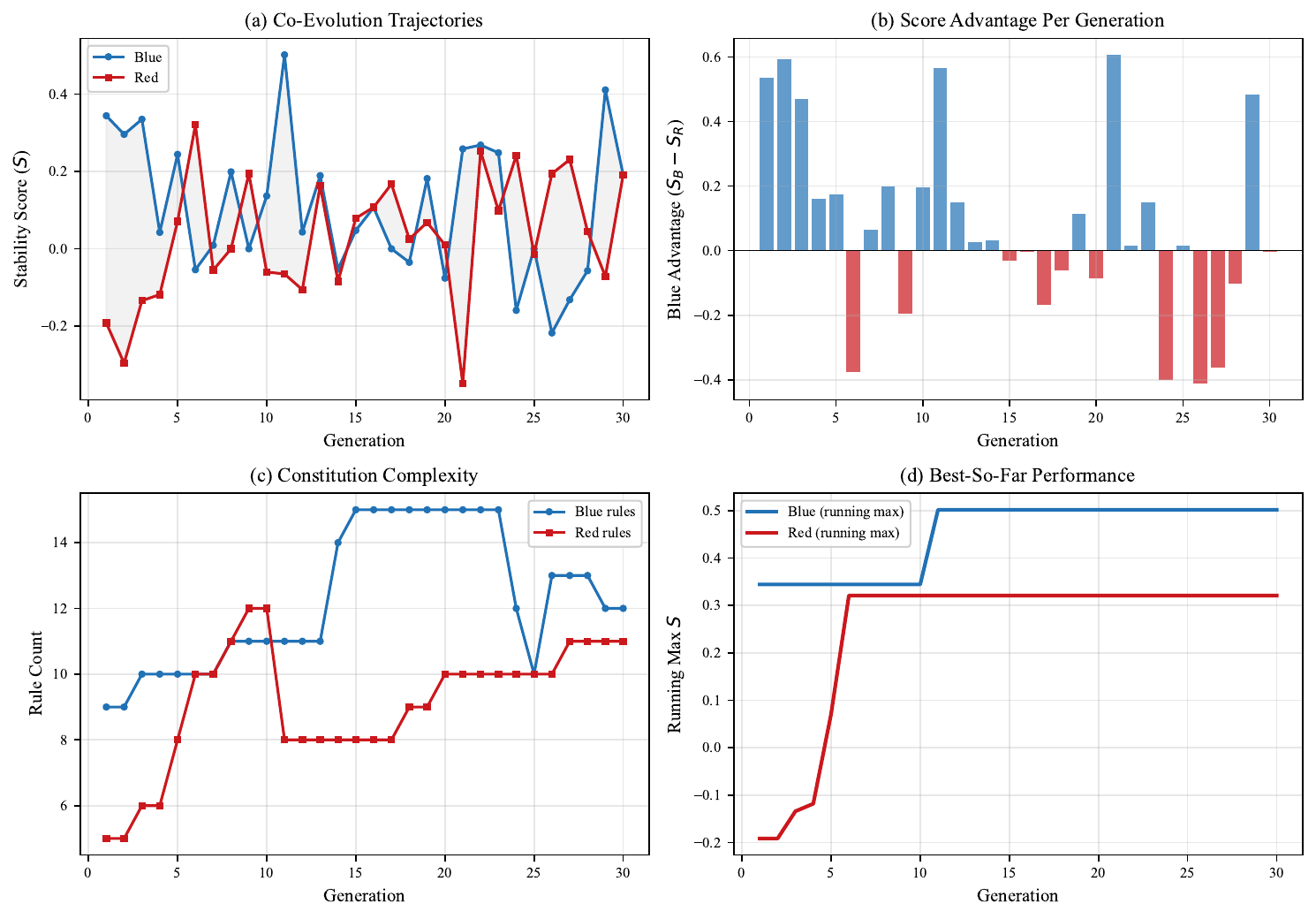}
    \caption{Full co-evolution arms race (Experiment~2) under score-advantage
    fitness.
    \textbf{(a)}~Stability scores for both factions across 30~generations;
    volatile early dynamics stabilize into near-parity oscillation.
    \textbf{(b)}~Blue advantage ($S_B{-}S_R$) per generation; mean near zero
    (dashed), consistent with zero-sum dynamics.
    \textbf{(c)}~Constitution complexity (rule count) over generations; both
    factions expand their rule sets as they adapt to each other.
    \textbf{(d)}~Best-so-far performance: running maximum $S$ for each
    faction, showing monotone improvement in the best discovered
    constitution despite per-generation volatility.}
    \label{fig:arms_race}
\end{figure}

\begin{table}[h]
\centering
\begin{minipage}{0.48\textwidth}
\caption{Exp~3, full information.}
\label{tab:exp3-full}
\centering
\begin{tabular}{ccc}
\toprule
\textbf{Gen} & \textbf{$S_B$} & \textbf{Red adv.} \\
\midrule
1  & $+0.267$ & $-0.225$ \\
10 & $+0.317$ & $-0.034$ \\
20 & $+0.062$ & $-0.033$ \\
25 & $+0.187$ & $+0.256$ \\
30 & $-0.102$ & $+0.267$ \\
\bottomrule
\end{tabular}
\end{minipage}\hfill
\begin{minipage}{0.48\textwidth}
\caption{Exp~3, asymmetric information (Red sees both factions, Blue sees
own team only).}
\label{tab:exp3-asym}
\centering
\begin{tabular}{ccc}
\toprule
\textbf{Gen} & \textbf{$S_B$} & \textbf{Red adv.} \\
\midrule
1  & $+0.339$ & $+0.264$ \\
10 & $+0.154$ & $-0.038$ \\
20 & $+0.000$ & $+0.503$ \\
25 & $+0.000$ & $+0.111$ \\
30 & $-0.138$ & $+0.415$ \\
\bottomrule
\end{tabular}
\end{minipage}
\end{table}

\section{Diagnostic Run: Independent Scoring Does Not Produce Adversarial Dynamics}
\label{app:preliminary}

This appendix documents the diagnostic argument in Section~\ref{sec:protocol}:
that independent per-faction scoring does not impose adversarial selection
pressure on Red.

\paragraph{Setup.} Blue fixed at $C^*$ from~\cite{kumar2026evolving}, Red
evolving from a zero-sum seed, 30~generations, $K{=}2$ seeds, fitness target
$S_R$ (per-faction independent scoring).

\paragraph{Statistics.} Pearson correlation between $S_B$ and $S_R$ across
30~generations: $+0.088$. Sum $S_B{+}S_R$ varied over $[1.09, 1.31]$ with
standard deviation $0.055$. Red's $S_R$ rose from a starting $0.483$ to a
maximum of $0.602$ at generation~7 and plateaued in $[0.49, 0.60]$ thereafter.
Blue's $S_B$ stayed in $[0.65, 0.73]$ for all 30~generations.

\paragraph{Inference.} A zero-sum game would imply negative correlation
between the factions' scores and approximately constant sum. Neither holds.
Under independent scoring, Red's evolutionary search has no gradient toward
strategies that degrade Blue: $S_R$ is a function only of Red-side quantities.
Red therefore evolves productive self-building behavior --- gathering
resources for its own projects --- rather than adversarial behavior.
Replacing the search target with score-advantage $S_R{-}S_B$
(Section~\ref{sec:protocol}) couples the factions' outcomes and produces
the qualitatively different dynamics reported in
Sections~\ref{sec:exp1}--\ref{sec:exp3}.

\paragraph{Note on the LLM-mutation syntax-error pathology.} Two conditions
in this paper (the own-team variant of Exp~3, and gens~2--10 of the PGG
pure-adversary run) returned $S_B{=}S_R{=}0$ from a recoverable failure
mode of the LLM mutation operator. The operator periodically produces
Python with syntax errors; at $K{=}2$ seeds, these broken constitutions
occasionally win the OpenEvolve database round. The orchestrator we used
at the time did not validate constitutions before saving them as the
next-generation seed; once a broken constitution became the parent for
further mutations, every child also failed to load and scored 0, poisoning
the lineage. In the PGG pure-adversary run, the OpenEvolve database recovered
from a valid earlier seed at generation~11. In the own-team Exp~3 condition,
the run completed before recovery. We have since added syntax validation
before checkpointing, but the affected runs are not recoverable without
reruns, which are planned for the camera-ready version.

\section{Constitution Specifications}
\label{app:constitutions}

\subsection{Grid-World: Initial Blue Constitution ($C^*$ Cooperative Seed,
from~\cite{kumar2026evolving})}
\begin{enumerate}
    \item \textbf{Deposit First} (Priority 1): If carrying any resource your team needs, deposit immediately.
    \item \textbf{Survival Focus} (Priority 2): Keep deposit count above elimination threshold.
    \item \textbf{Gather \& Deposit} (Priority 3): Collect needed resources when empty.
    \item \textbf{Dynamic Targeting} (Priority 4): Move toward largest team deficit.
    \item \textbf{Share Resources} (Priority 5): Transfer surplus to nearby teammates.
    \item \textbf{Report Rich Cluster} (Priority 6): Broadcast only for 2+ resources.
    \item \textbf{Conflict Avoidance} (Priority 7): No aggression unless attacked.
\end{enumerate}

\subsection{Grid-World: Initial Red Constitution (Zero-Sum Seed)}
\begin{enumerate}
    \item \textbf{Sabotage Opponents} (Priority 1): Attack or steal from opposing team.
    \item \textbf{Hoard Resources} (Priority 2): Gather but delay depositing.
    \item \textbf{Self-Preservation} (Priority 3): Focus on personal survival.
\end{enumerate}

\subsection{PGG: Initial Blue Constitution (Cooperator Seed)}
\begin{enumerate}
    \item \textbf{Contribute Generously} (Priority 1): Contribute at least 60\% each round.
    \item \textbf{Conditional Cooperation} (Priority 2): Reduce if group average drops below 30\%.
    \item \textbf{Punish Free-Riders} (Priority 3): Punish agents contributing $<$10\% for 2+ rounds.
    \item \textbf{Communicate Norms} (Priority 4): Broadcast cooperation expectations.
    \item \textbf{Maintain Fairness} (Priority 5): Match or exceed group average.
\end{enumerate}

\subsection{PGG: Initial Red Constitution (Free-Rider Seed)}
\begin{enumerate}
    \item \textbf{Minimise Contribution} (Priority 1): Contribute at most 15\%.
    \item \textbf{Exploit Cooperators} (Priority 2): Contribute 0\% when others contribute highly.
    \item \textbf{Avoid Detection} (Priority 3): Occasionally contribute 20--30\% to avoid punishment.
    \item \textbf{Deceptive Signalling} (Priority 4): Announce high contributions, contribute low.
    \item \textbf{Late-Round Defection} (Priority 5): Cooperate early, defect after round 3.
\end{enumerate}

\section{Hyperparameters}
\label{app:hyperparameters}

\begin{table}[h]
\centering
\caption{Evolution hyperparameters (both environments).}
\begin{tabular}{lc}
\toprule
\textbf{Parameter} & \textbf{Value} \\
\midrule
Generations & 30 \\
Population size (per generation) & 6 \\
Islands per generation & 1 \\
Elite selection ratio & 0.4 \\
Exploitation ratio & 0.4 \\
Exploration ratio & 0.2 \\
LLM model & GPT-OSS-120B \\
Temperature & 1.0 \\
Evaluation seeds per constitution ($K$) & 2 (default); 5 (Ablation~1, Section~\ref{sec:exp5}) \\
Grid-world: turns per simulation & 40 \\
Grid-world: grid size & $8 \times 8$ \\
PGG: rounds per game & 20 \\
PGG: pool multiplier & 1.5 (main); $\{1.2, 2.0, 3.0\}$ (ablation) \\
PGG: initial endowment & 100 \\
PGG: punishment cost ratio & 0.3 \\
\bottomrule
\end{tabular}
\end{table}

\section{Reproducibility}
\label{app:reproducibility}

Code, configurations, and initial constitutions will be released upon
publication. Experiments use GPT-OSS-120B~\citep{openai2025gptoss} via
OpenRouter with temperature 1.0. Random seeds are fixed at $42 + 17k$ for
seed index $k$.

\end{document}